\documentclass[iop]{emulateapj}
\usepackage[pass,letterpaper]{geometry}
\usepackage{hyperref}

\journalinfo{ApJL in press}
\submitted{Received 2014 November 14; accepted 2014 December 31}

\shorttitle{A Continuum of Small Planets}
\shortauthors{Schlaufman}

\begin{document}

\title{A CONTINUUM OF SMALL PLANET FORMATION BETWEEN 1 AND 4 EARTH RADII}

%

\author{Kevin C.\ Schlaufman\altaffilmark{1,2}}
\affil{
$^1$ Kavli Institute for Astrophysics and Space Research, Massachusetts Institute of Technology, Cambridge, MA 02139, USA; kschlauf@mit.edu}

\altaffiltext{2}{Kavli Fellow.}

\begin{abstract}

\noindent
It has long been known that stars with high metallicity are more
likely to host giant planets than stars with low metallicity.  Yet the
connection between host star metallicity and the properties of small
planets is only just beginning to be investigated.  It has recently
been argued that the metallicity distribution of stars with exoplanet
candidates identified by \textit{Kepler} provides evidence for three
distinct clusters of exoplanets, distinguished by planet radius
boundaries at $1.7\,R_{\oplus}$ and $3.9\,R_{\oplus}$.  This would
suggest that there are three distinct planet formation pathways for
super-Earths, mini-Neptunes, and giant planets.  However, as I show
through three independent analyses, there is actually no evidence for
the proposed radius boundary at $1.7\,R_{\oplus}$.  On the other hand,
a more rigorous calculation demonstrates that a single, continuous
relationship between planet radius and metallicity is a better fit to
the data.  The planet radius and metallicity data therefore provides no
evidence for distinct categories of small planets.  This suggests that
the planet formation process in a typical protoplanetary disk produces
a continuum of planet sizes between $1\,R_{\oplus}$ and $4\,R_{\oplus}$.
As a result, the currently available planet radius and metallicity data
for solar-metallicity F and G stars give no reason to expect that the
amount of solid material in a protoplanetary disk determines whether
super-Earths or mini-Neptunes are formed.

\end{abstract}


\section{Introduction}

The probability that a giant planet orbits a star is a steeply rising
function of the host star's metallicity \citep[e.g.,][]{san04,fis05}.
This observation is the key piece of evidence that the giant planets
identified by the radial velocity and transit techniques form through core
accretion and not through gravitational instability.  This observation
is perhaps the most important constraint placed on models of planet
formation since the discovery of the first exoplanets.

The connection between stellar metallicity and the presence of small
planets is less clear.  The Neptune-mass planets discovered by radial
velocity surveys do not appear to preferentially orbit metal-rich FGK
stars \citep[e.g.,][]{sou08,may11}.  While \textit{Kepler} has discovered
a large number of small exoplanet candidates (planets from here), it
has not yet settled the issue.  \citet{sch11} showed that while the
giant planets discovered by \textit{Kepler} orbit metal-rich stars, the
small planets discovered around F and G stars did not appear to prefer
metal-rich stars.  This observation was later confirmed by \citet{buc12}.

Recently, \citet{buc14} (B14 from here) argued that the observed
distribution of metallicity in a sample of more than 400 \textit{Kepler}
planet host stars revealed three distinct clusters of exoplanets:
terrestrial planets with planet radius $R_{p}\lesssim 1.7\,R_{\oplus}$,
``gas dwarf" planets with $1.7\,R_{\oplus}\lesssim R_{p}\lesssim
3.9\,R_{\oplus}$, and ice or gas giants with $R_{p} \gtrsim
3.9\,R_{\oplus}$.  They suggested that these three populations formed
via distinct planet formation channels.

To reach that conclusion, B14 repeatedly split their sample of planet
host star metallicity measurements into small-planet and large-planet
subsamples for different choices of the radius boundary dividing
the two subsamples.  They calculated the $p$-value from a two-sample
Kolmogorov--Smirnov test on the two subsamples as a function of planet
radius and identified local minima $p$-values.  They saved the radii
at which the local minima occurred.  To account for measurement
uncertainties, they repeated this process $10^{6}$ times, sampling
the planet radius and host star metallicity from their uncertainty
distributions on each iteration.  They identified a distinct $p$-value
minimum at $R_{p}=1.7\,R_{\oplus}$ and argued that it represents a
boundary between terrestrial and ``gas dwarf" planets.  This approach
is inappropriate because it performs a large number of hypothesis tests
on the same data set without correcting the test thresholds to account
for the large number of tests.  That strategy is known to produce a
high false-discovery rate \citep[e.g.,][]{dun59,dun61}.  Moreover, the
B14 technique creates a sequence of $p$-values at many split points for
data subject to measurement uncertainty.  Consequently, before attaching
any significance to features in that sequence of $p$-values, it is also
critical to ensure that the $p$-values that result from the Monte Carlo
simulation accurately represent the $p$-value measurement uncertainties
that result from uncertainties in the input sample.

There are at least four more problems with the analysis presented in B14.
First, B14 overlooked the effect of planet radius uncertainty due to
transit depth uncertainty.  Second, their approach used an asymptotically
inconsistent estimator of the average $p$-value at each split point
in the presence of observational uncertainty.  Third, their analysis
is subject to the multiple comparisons problem, which reduces the
significance of their observation by a large amount.  Fourth, while B14
assert that local minima in a plot of $p$-value as a function of split
radius indicate transitions between distinct clusters of exoplanets,
this is not necessarily so.  I describe my sample selection in Section 2,
I detail each issue with the B14 calculation in Section 3, I outline a
more rigorous way to investigate the issue in Section 4, and I discuss
the implications and my conclusion in Section 5.

\section{Sample Construction}

\begin{figure*}
\plotone{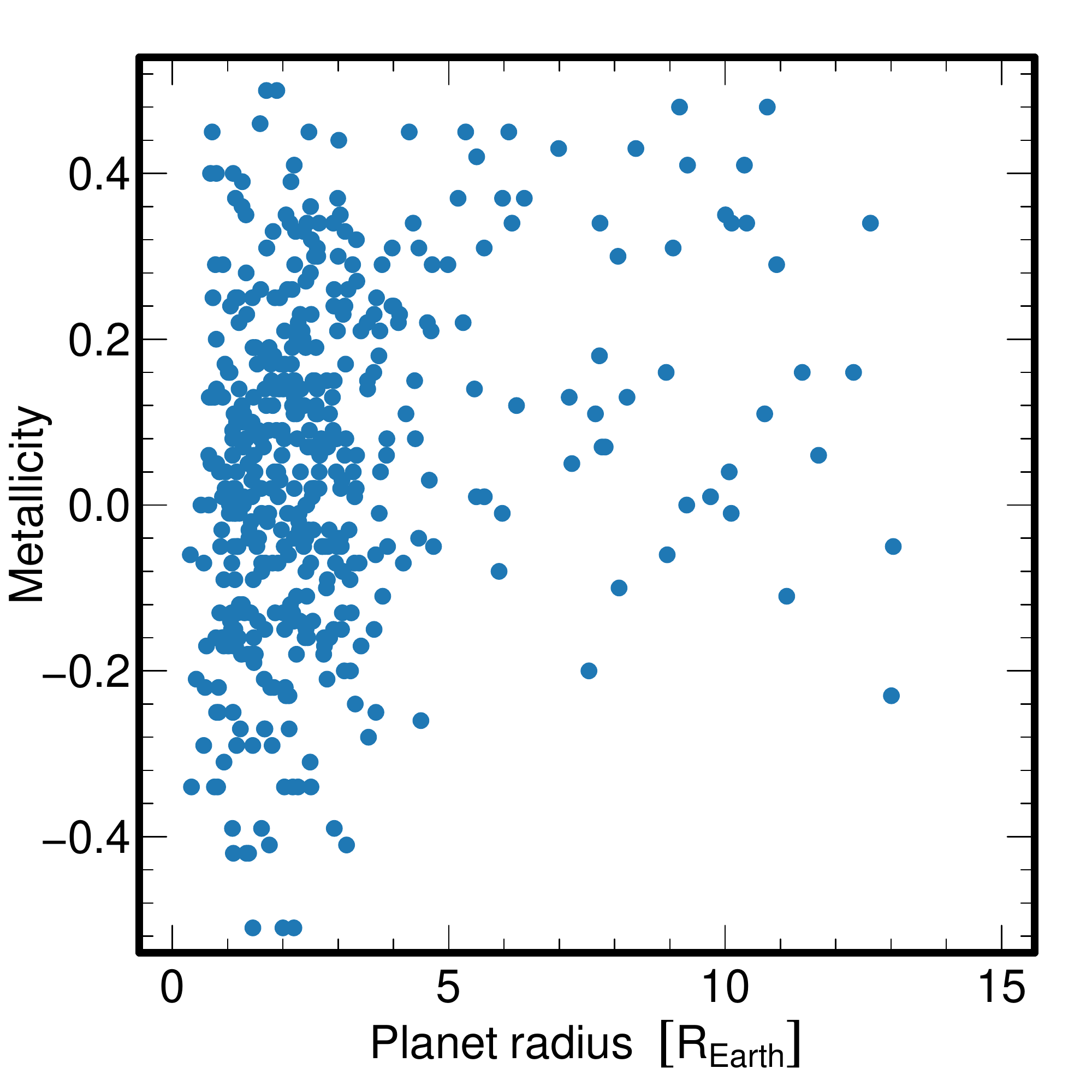}
\caption{Planet radius $R_{p}$ vs. host star metallicity.\label{fig01}}
\end{figure*}

I use the planet host star data from B14.  Those data include
$T_{\mathrm{eff}}, \log{g}$, [M/H], $M_{\ast}$, $R_{\ast}$, and
their associated uncertainties.  B14 did not include in their planet
radius uncertainties the effect of uncertainties in transit depth,
even though transit depth uncertainties are more important than the
host star radius uncertainties in 25\% of the sample.  As a result,
I supplement the B14 data with the latest \textit{Kepler} object of
interest period and $R_{p}/R_{\ast}$ estimates from the \textit{Kepler}
CasJobs database\footnote{\url{http://mastweb.stsci.edu/kplrcasjobs/}}
hosted by the Mikulski Archive for Space Telescopes.  I then recompute
planet radii from the B14 stellar radii and the updated transit depths.
Following B14, I remove from the sample all planets smaller than
$3\,R_{\oplus}$ subject to strong stellar irradiation (i.e., $F_{\nu}
> 5 \times 10^{5}$ J s$^{-1}$ m$^{-2}$), as these planets may have lost
a significant fraction of their initial atmospheres.  I plot these data
in Figure~\ref{fig01}.

\section{Issues With the Buchhave et al.\ (2014) Calculation}	

\subsection{An Asymptotically Inconsistent p-value Estimator}

An asymptotically inconsistent estimator of a parameter does not
converge to the true value of the parameter in the large-sample limit.
One problem with the B14 analysis is that they used an asymptotically
inconsistent estimator of the $p$-value averaged over planet radius
measurement uncertainty in their Monte Carlo simulation.  The $p$-value
measurements depend on the planet radius measurements, which are subject
to measurement uncertainty in the inferred stellar radii $R_{\ast}$
and the measured ratios $R_{p}/R_{\ast}$.  The true $p$-values in the
absence of uncertainty cannot be measured directly.  Instead, one can
only measure $p'$
\begin{eqnarray}
p' & = & p + N\left(0,\sigma^2\right),
\end{eqnarray}
\noindent
where $p$ is the true $p$-value and $N(0,\sigma^2)$ is due to measurement
uncertainties in $R_{\ast}$ and $R_{p}/R_{\ast}$.  Repeatedly calculating
$p'$ after perturbing each planet radius due to the uncertainties in
$R_{\ast}$ and $R_{p}/R_{\ast}$ and averaging the result will provide
an asymptotically consistent estimate of $p$ by the central limit theorem
\begin{eqnarray}
E\left[p'\right]&=&E\left[p+N(0,\sigma^2)\right],\\
E\left[p'\right]&=&E\left[p\right]+E\left[N(0,\sigma^2)\right],\\
\frac{1}{n}\sum_{i=1}^{n} p'&=& \frac{1}{n}\sum_{i=1}^{n} p+\frac{1}{n}\sum_{i=1}^{n}N(0,\sigma^2),\\
\overline{p'}&=& \overline{p}+0\Rightarrow\overline{p'}=\overline{p}.
\end{eqnarray}

B14 never averaged the $p$-value produced for each split point over
all iterations.  Instead, after each iteration of their Monte Carlo
simulation they identified the local $p$-value minima and saved them.
After completing 10$^6$ Monte Carlo iterations, they determined the mean
radii at which local $p$-value minima occurred by averaging over the
individual radii calculated on each iteration.  In other words, they
applied the nonlinear function $f$ that takes a sequence of $p$-values
and identifies the radii of local $p$-value minima before averaging over
all iterations to identify the mean radii at which $p$-value minima occur.
The central limit theorem does not apply in this case, as
\begin{eqnarray}
f(p')&=&f\left(p+N(0,\sigma^2)\right),\\
E\left[f(p')\right]&=&E\left[f\left(p+N(0,\sigma^2)\right)\right],\\
\frac{1}{n}\sum_{i=1}^{n}f(p')&=&\frac{1}{n}\sum_{i=1}^{n}f\left(p+N(0,\sigma^2)\right),\\
\overline{f(p')}&=&\overline{f\left(p+N(0,\sigma^2)\right)}\not\Rightarrow\overline{f(p')}=\overline{f(p)}.
\end{eqnarray}
\noindent
As a result, the $p$-values in their Figure 1 improperly account for
measurement uncertainty and are asymptotically inconsistent with the
true $p$-values absent measurement uncertainty.

\begin{figure*}
\plotone{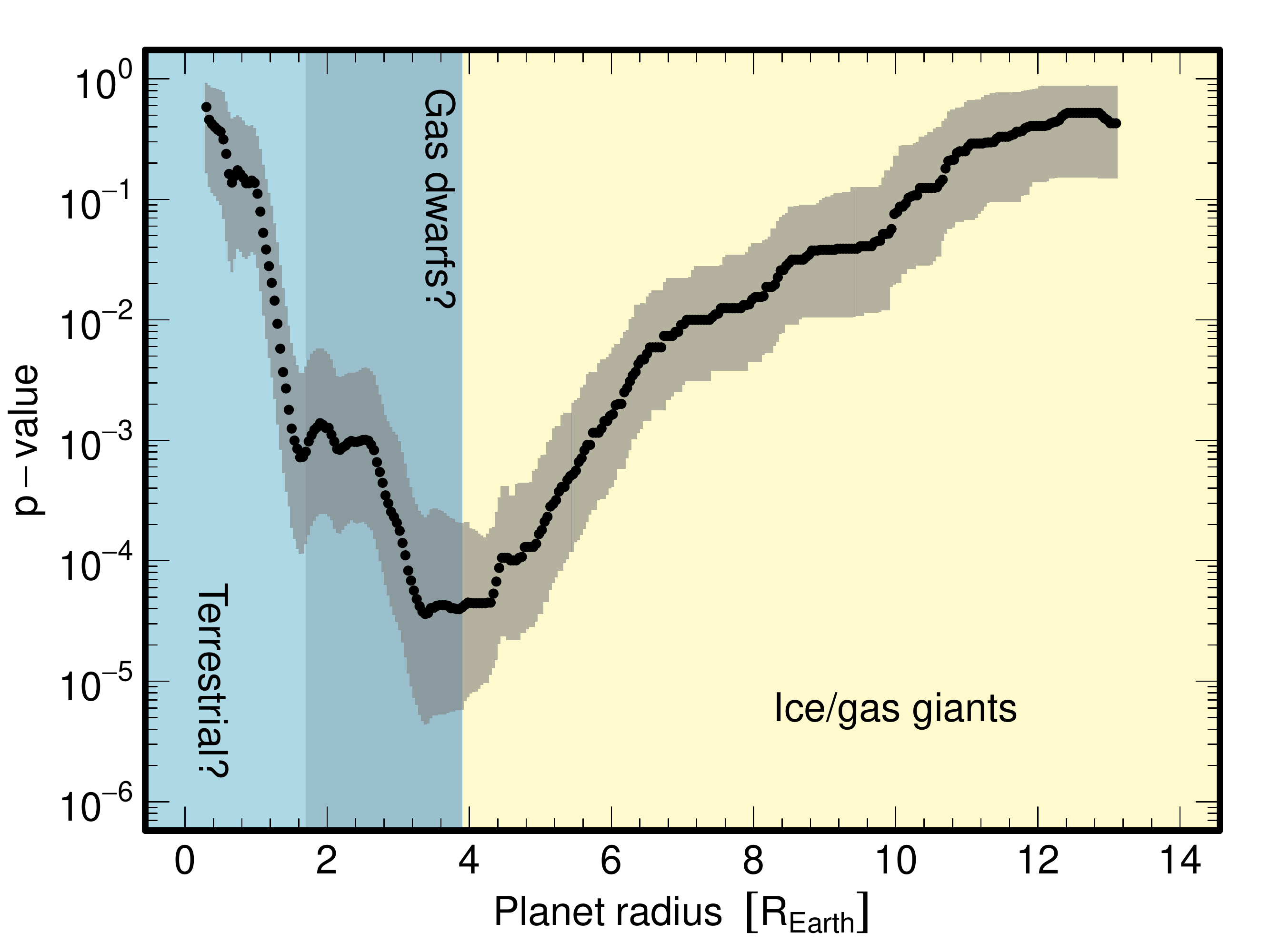}
\caption{Mean $p$-value as a function of planet radius.  I split
the sample into small-planet and large-planet subsamples at 321
split points from $0.3\,R_{\oplus}$ to $13.1\,R_{\oplus}$ in steps
of $0.04\,R_{\oplus}$ and compute the $p$-value from a two-sample
Kolmogorov--Smirnov tests on the metallicity distributions of
both subsamples.  I repeat this process 10$^{5}$ times.  The black
points are mean $p$-values averaged over the uncertainties in host
star radius $R_{\ast}$ and transit depth $(R_{p}/R_{\ast})^2$.
I indicate the uncertainty at each radius as a semi-transparent gray
rectangle with height given by the uncertainty in the $p$-value and
width $0.02\,R_{\oplus}$.  After accounting for the uncertainties in
$R_{\ast}$ and $(R_{p}/R_{\ast})^2$, the $p$-values do not support the
idea of a qualitative difference between planets with radii above or
below $1.7\,R_{\oplus}$.\label{fig02}}
\end{figure*}

To address that problem, I first generate $10^{5}$ realizations of each
planet radius from the distributions that result from the propagation of
measurement uncertainties in $R_{\ast}$ and $R_{p}/R_{\ast}$.  I split
the metallicity data into small-planet and large-planet subsamples at 321
split points from 0.3 to $13.1~R_{\oplus}$ in steps of $0.04~R_{\oplus}$
and compute the $p$-value from a two-sample Kolmogorov--Smirnov test.
I save the resulting $p$-value for each split point and repeat this
process $10^{5}$ times.  At the end of the calculation, I average the
$p$-values for each split point.  I plot the result in Figure~\ref{fig02}.
The apparent local minimum in the $p$-value distribution identified by
B14 at $R_{p}=1.7\,R_{\oplus}$ is not present.

\subsection{The Multiple Comparisons Problem}

Another issue involves the multiple comparisons problem.  The multiple
comparisons problem occurs in statistical analyses when the same
data is both used to select a model and estimate its parameters
\citep[e.g.,][]{ben10}.  It frequently leads to the underestimate of
the uncertainty of the model parameters.  In this case, B14 used the
same metallicity data to both identify the planet radius boundaries that
separated the three distinct clusters and to estimate the mean metallicity
and associated uncertainty for each cluster.  Since they used their data
both to set the boundaries and determine the mean metallicities for each
region, their analysis is subject to the multiple comparisons problem.

One way to correct for this problem is to use independent data sets, one
to select the model and another to fit the model parameters.  In this
case, the correct approach is to split the metallicity data in half.
The first half should be used to identify the planet radius boundaries
that separate the three distinct clusters of exoplanets.  The second half
should then be used to infer the average metallicity of each proposed
cluster.  This process can be repeated a large number of times with
different randomly selected subsamples.  Consequently, on each iteration
of a Monte Carlo simulation, I split the data set described in Section
2 in half.  I follow the approach of B14 and identify $p$-value minima
at $R_{p} < 2\,R_{\oplus}$ and $2\,R_{\oplus} < R_{p} < 4\,R_{\oplus}$.
I use those planet radii as the boundaries of each exoplanet cluster
and use the second half of the metallicity data to compute the mean
metallicity of each cluster.  I repeat this process $10^{5}$ times.
I find that the difference between the mean metallicities for the
terrestrial and ``gas dwarf" regions is only 0.7$\sigma$---much lower
than the 3.1$\sigma$ offset reported by B14.

Metal-poor stars are smaller than metal-rich stars, so a bias toward
finding small planets around metal-poor stars in a transit-depth-limited
survey is a systematic effect that will further decrease the significance
of this offset \citep{gai13}.  The fact that the mean metallicities of the
stars on either side of the claimed transition at $R_{p}=1.7\,R_{\oplus}$
are indistinguishable contradicts the B14 interpretation of the transition
as evidence of different planet formation pathways.

\subsection{Do Local p-value Minima Indicate Distinct Exoplanet Regimes?}

\begin{figure*}
\plotone{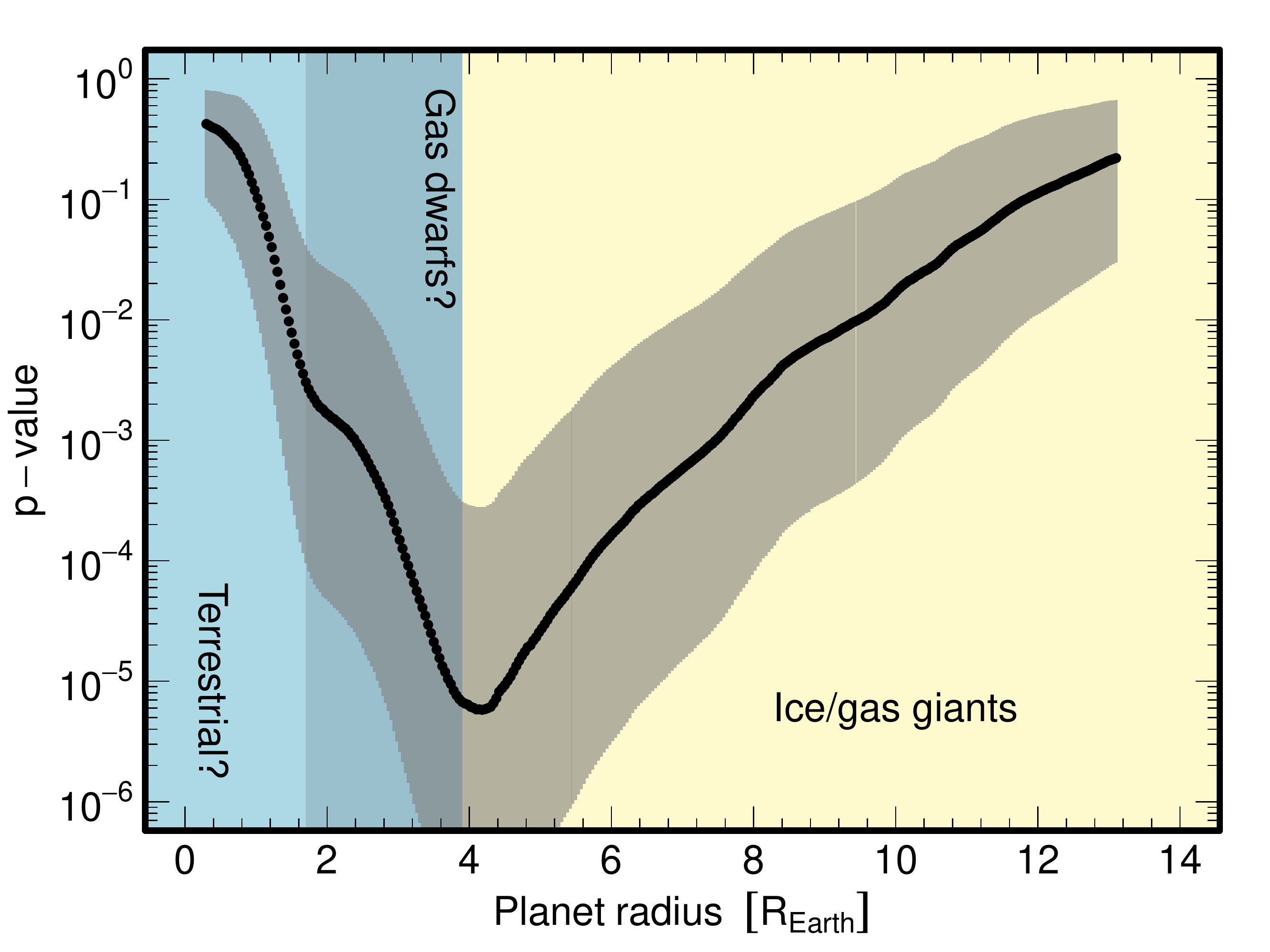}
\caption{Mean $p$-value as a function of planet radius in the
scenario advocated by B14 for three distinct planet clusters
separated at $1.7\,R_{\oplus}$ and $3.9\,R_{\oplus}$ with metallicity
distributions $N(0.00,0.20^2)$, $N(0.05,0.19^2)$, and $N(0.18,0.19^2)$.
If planet radius boundaries at $1.7\,R_{\oplus}$ and $3.9\,R_{\oplus}$
do separate the exoplanet population into three clusters with unique
metallicity distributions, then that difference would manifest
itself as a non-continuous first derivative---a ``kink"---at
$1.7\,R_{\oplus}$.  Even though three unique metallicity distribution
were imposed by construction in this case, there is no $p$-value minimum
at $1.7\,R_{\oplus}$.  Consequently, even if there were three distinct
clusters of exoplanets, each with a unique metallicity distribution,
the analysis described in B14 would not be able to identify
them by $p$-value minima.\label{fig03}}
\end{figure*}

While B14 argue that local minima in a plot of split radius versus
$p$-value indicate transitions between distinct exoplanet clusters,
this is not always the case.  To demonstrate this, I use the same Monte
Carlo simulation described in Section 3.1.  However, instead of using
the observed metallicities, on each iteration I randomly sample the
metallicities of stars hosting planets with $R_{p}\leq1.7\,R_{\oplus}$,
$1.7\,R_{\oplus}<R_{p}\leq3.9\,R_{\oplus}$, and $R_{p}\geq3.9\,R_{\oplus}$
from their observed distributions.  Those distributions are
$N(0.00,0.20^2)$, $N(0.05,0.19^2)$, and $N(0.18,0.19^2)$.  I plot
the result in Figure~\ref{fig03}.  Despite the fact that distinct
metallicity distributions were imposed on each planet cluster, there is
no local minimum in the $p$-value distribution at the boundary between
the terrestrial and ``gas dwarf" planets.  The inability of the B14
technique to identify a metallicity boundary imposed by construction as
a local $p$-value minimum implies that the technique is not sensitive
to subtle features in the metallicity distribution.

\section{A More Rigorous Approach}

A better way to identify the number of subpopulations required by the
planet radius and host star metallicity data is to compare statistical
models with varying numbers of components, then identify the model
that has the minimum number of parameters yet the maximum likelihood of
producing the observed data.  I consider two classes of models.  First,
I fit single-population linear models of the form
\begin{eqnarray}
\mathrm{[M/H]} & = & a_{0} + \sum_{j=1}^{m} a_{j} R_{p}^{j} + \epsilon,
\end{eqnarray}
where $\epsilon$ is the standard uncertainty term in the regression
equation.  Second, I fit finite Gaussian mixture models with varying
numbers of subpopulations of the form
\begin{eqnarray}
\prod_{i=1}^{n} \sum_{j=1}^{m} w_{j} N_{j}(\mathbf{x}_{i}|\mathbf{\mu}_{j},\mathbf{\Sigma}_{j}),
\end{eqnarray}
where $\mathbf{x}$ is the data, $m$ is the number of components in
the model, the $w_{j}$ are weights such that $\sum_{j=1}^{m} w_{j} =
1$, and each $N_{j}$ is a two-dimensional Gaussian component of the
overall density
\begin{eqnarray}
N_{j}(\mathbf{x}|\mathbf{\mu}_{j},\mathbf{\Sigma}_{j}) & = & \frac{1}{2\pi |\mathbf{\Sigma}_{j}|^{1/2}} \times \nonumber\\
& & \exp{\left\{-\frac{1}{2} (\mathbf{x}_{i}-\mathbf{\mu}_{j})^T \mathbf{\Sigma}_{j}^{-1} (\mathbf{x}_{i}-\mathbf{\mu}_{j}) \right\}}.
\end{eqnarray}
Here $\mathbf{\mu}_{j}$ and $\mathbf{\Sigma}_{j}$ are
the mean and covariance of each of the $m$ components
of the model.  I fit the Gaussian mixture models using the
\texttt{mclust}\footnote{\url{http://www.stat.washington.edu/mclust/}}
package in \texttt{R}\footnote{\url{http://www.R-project.org/}}
\citep{fra02,fra12,r14}.

\begin{figure*}
\plottwo{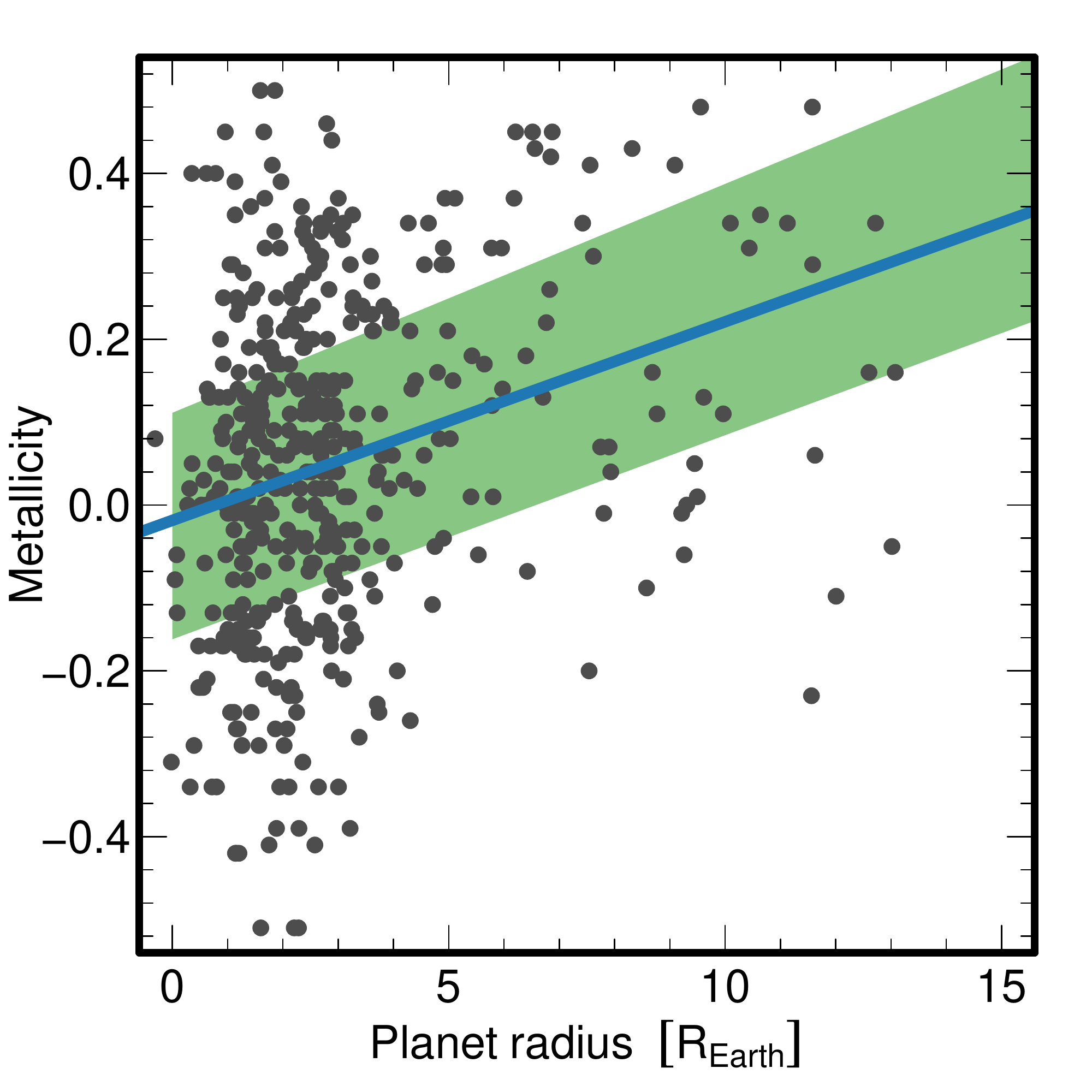}{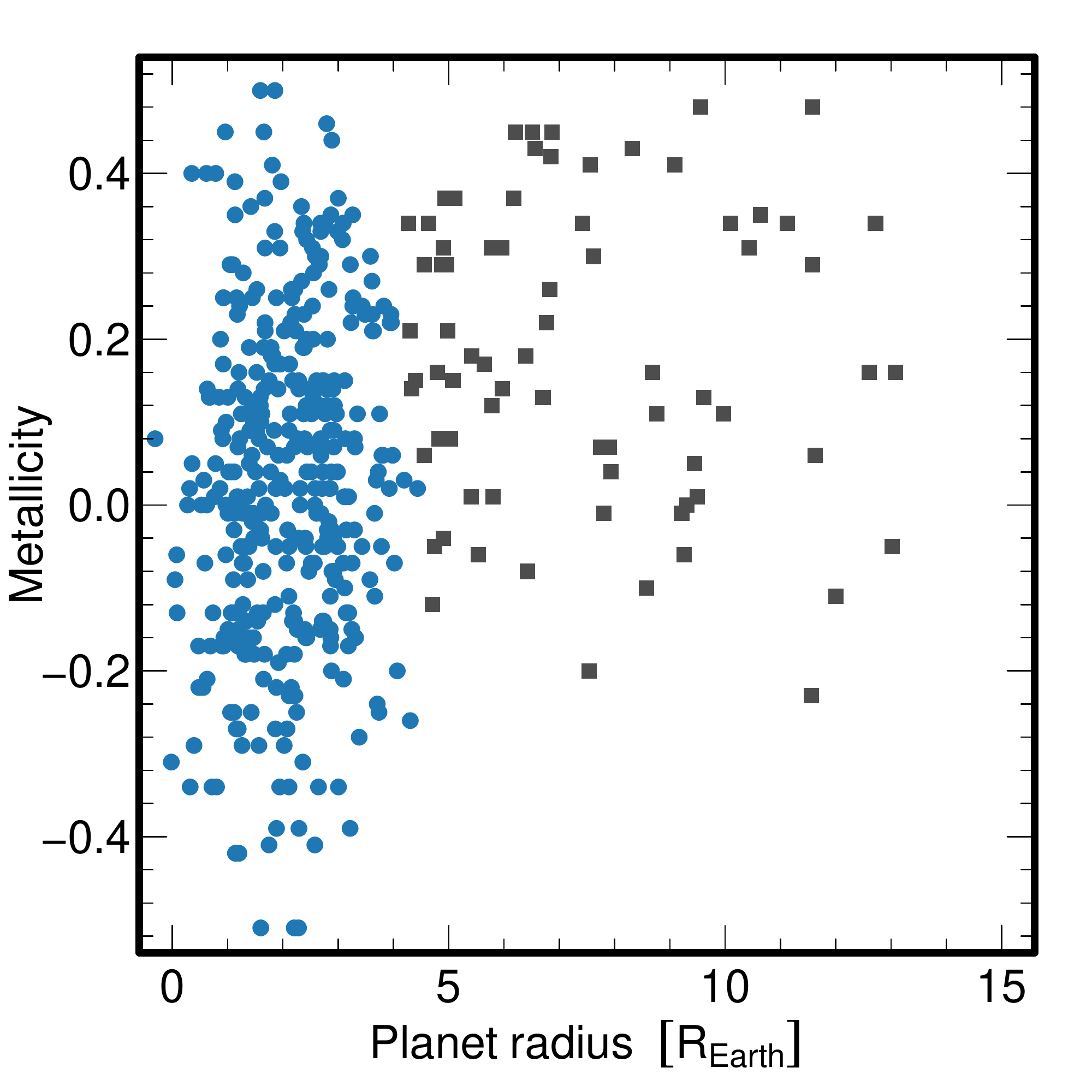}
\caption{Two different models for the relationship between $R_{p}$
and metallicity.  Left: a one-component, linear relationship between
$R_{p}$ and metallicity.  The blue line shows the best-fit model, while
the green-shaded region shows the 25\% and 75\% quantile regression
bands computed using the \texttt{quantreg} package \citep{koe13}.
Right: a Gaussian mixture model with two components.  Planets plotted
as blue circles are assigned to one component, while gray squares are
assigned to the other.  The divide between the two populations occurs at
$R_{p}\approx4\,R_{\oplus}$.  While the two-component Gaussian mixture
model is favored over mixture models with one to seven components,
the one-component linear model is favored by the Akaike information
criterion (AIC) over any of the mixture models.  For that reason,
a one-component smooth model is currently the best match to the
\textit{Kepler} planet radius and metallicity data for F and G stars
presented in B14.\label{fig04}}
\end{figure*}

To account for the observational uncertainties, I use a Monte Carlo
simulation.  I sample the planet radii from the distributions that
result from the propagation of measurement uncertainties in $R_{\ast}$
and $R_{p}/R_{\ast}$ and directly use the measured metallicities
(since the uncertainty in [M/H] is already reflected in the uncertainty
in $R_{\ast}$).  On each iteration, I fit linear models of the form of
Equation (10) for $m=1,2, \ldots,5$ and Gaussian mixture models $m=1,2,
\ldots,7$.  I choose both the best linear and Gaussian mixture models
using the Bayesian information criterion \citep[BIC;][]{sch78}, then use
the Akaike information criterion \citep[AIC;][]{aka74} to choose between
the favored linear and Gaussian mixture models.  I repeat this process
$10^{3}$ times.  In all cases, the best linear model is preferred.
For the linear model, the $m=1$ model is favored 76.4\% of the time,
the $m=2$ model is favored 22.3\% of the time, while a higher-order
model is favored 1.3\% of the time.  While the Gaussian mixture model is
disfavored relative to the linear model, the two-component model is the
best of the mixture models: the two-component model is preferred on 92.8\%
of the iterations, while the three-component model is preferred 7.2\%
of the time.  I plot representative examples of the BIC-selected models
from one iteration of my Monte Carlo simulation in Figure~\ref{fig04}.

\section{Discussion and Conclusion}

The performance of a large number of tests on the same data set without
correcting the test thresholds, the use of an asymptotically inconsistent
estimator of the $p$-value in the presence of measurement uncertainty, the
oversight of the multiple comparison problem, or the inability of the B14
technique to identify an imposed metallicity effect as a $p$-value minimum
are all sufficient reasons to be skeptical of the claimed transition at
$R_{p}=1.7\,R_{\oplus}$.  The problems with the B14 analysis technique
cannot be mitigated by examining a larger or independent data set---they
are inherent in the analysis technique itself.  Instead, the analysis in
Section 4 shows that a smooth, one-component linear model is a better
fit to the data than any multi-component model.  If a multi-component
model is used, then the two-component model is consistently a better
fit than the three-component model.  As a result, the planet radius
and metallicity data for the \textit{Kepler} F and G star planet hosts
does not support the idea of multiple types of small planets.  Instead,
a continuum of planet sizes between $1\,R_{\oplus}$ and $4\,R_{\oplus}$
are likely formed independent of the amount of solids present.

While observational evidence suggests that most planets larger
than about $2\,R_{\oplus}$ have significant hydrogen atmospheres
\citep[e.g.,][]{mar14}, Kepler-10c is an exception with a radius of
$2.35\,R_{\oplus}$ and a density of 7.1 g cm$^{-3}$ \citep{dum14}.
Likewise, smaller planets probably have a wide range of atmospheric
properties \citep[e.g.,][]{rog14,wol14}.  Moreover, a wide range of
densities can be present even in the same system, with Kepler-36 the
best example \citep{car12}.  For these reasons, near solar metallicity
it does not seem likely that the final masses or compositions of small
exoplanets are controlled primarily by the amount of solid material
present in their parent protoplanetary disks.

\acknowledgments
I thank Lars Buchhave, Andy Casey, Bryce Croll, David W.\ Latham,
Dimitar Sasselov, and Josh Winn.  I am especially grateful to the
referee Eric Feigelson for suggestions that substantially improved this
paper.  This research has made use of NASA's Astrophysics Data System
Bibliographic Services.  Some of the data presented in this paper were
obtained from the Mikulski Archive for Space Telescopes (MAST). STScI is
operated by the Association of Universities for Research in Astronomy,
Inc., under NASA contract NAS5-26555. Support for MAST for non-HST data
is provided by the NASA Office of Space Science via grant NNX13AC07G
and by other grants and contracts.  This paper includes data collected
by the \textit{Kepler} mission. Funding for the \textit{Kepler} mission
is provided by the NASA Science Mission directorate.  Support for this
work was provided by the MIT Kavli Institute for Astrophysics and Space
Research through a Kavli Postdoctoral Fellowship.

{\it Facilities:} \facility{Kepler}

\clearpage

\clearpage

\clearpage

\clearpage

\end{document}